\documentclass[prd,amsfonts,twocolumn,nofootinbib,showpacs]{revtex4}
\RequirePackage{graphicx}
\usepackage{enumerate}

\begin{document}
\title{Scale-dependent CMB asymmetry from primordial configuration}

\author{Kazunori Kohri}
\affiliation{Cosmophysics group, Theory Center, IPNS, KEK,  
and The Graduate University for Advanced Study (Sokendai), Tsukuba 305-0801, Japan}
\author{Chia-Min Lin}
\affiliation{Department of Physics, Chuo University, Bunkyo-ku, Tokyo
112, Japan}
\author{Tomohiro Matsuda}
\affiliation{Laboratory of Physics, Saitama Institute of Technology, Fukaya, Saitama 369-0293, Japan}

\begin{abstract}
We demonstrate that a topological defect can explain the hemispherical
 power asymmetry of the CMB.
The first point is that a defect configuration, which already exists
 prior to inflation, can source asymmetry of the CMB.
The second point is that modulation mechanisms, such as the curvaton and
other modulation mechanisms, can explain scale-dependence of the asymmetry. 
Using a simple analysis of the $\delta N$ formalism, we show models in
 which scale-dependent hemispherical power asymmetry is explained by
 primordial configuration of a defect.
\end{abstract}

\maketitle

\section{Introduction}
After the report of the detection of the hemispherical power asymmetry on the
CMB~\cite{WMAPCMBA, PlanckCMBA}, there have been attempts to look for
fingerprints of the non-standard inflationary physics in the anomaly.
If the anomaly is not a statistical artifact, it strongly indicates that
the single-field inflation models are not enough to explain the current
observations of the Universe~\cite{Hemisphere-original}.
Also, the anomaly seems to be suggesting that we are observing a
fingerprint of the pre-inflationary configuration~\cite{Hemisphere-original}.
If the dipolar modulation of the temperature $T(\vec{r})$ is generated
by the dipolar modulation of the curvature perturbation $\zeta$, a
parametrization of the bipolar asymmetry can be defined as 
\begin{equation}
\label{eq1}
\frac{{\cal P}^{1/2}_{\zeta}(k,\vec{r})}
{\left.{\cal P}^{1/2}_{\zeta}(k,\vec{r})\right|_\mathrm{iso}}=
1+\frac{A(k)(\vec{p}\cdot\vec{r})}{r_\mathrm{CMB}},
\end{equation}
where ${\left.{\cal P}_{\zeta}(k,\vec{r})\right|_\mathrm{iso}}$ is the
isotropic power spectrum and
$A(k)$ measures the asymmetry in the direction of the unit vector $\vec{p}$.
Here $r_\mathrm{CMB}$ is the comoving distance to the surface of the last scattering.

The effect of a large-scale enhancement of the spectrum of the curvature
perturbation~\cite{Lyth-book} has been investigated in Ref.\cite{GZ1978} and it has been
shown that the enhancement proportional to a linear function of
position ($z$) is not observable because of
the Doppler shift due to the induced peculiar velocity. 
If the enhancement has higher order corrections ($\sim z^n$), the second order term
($\sim z^2$) enhances the CMB quadrupole (Grishchuk-Zel'dovich effect),
which has not been observed yet. 
Then in Ref.\cite{Hemisphere-original}, a bound has been 
found for both inflationary scenario and 
curvaton model upon mild assumption of a function.
The observed CMB asymmetry has been explained by introducing additional
super-horizon size 
perturbation $\Delta\phi\sim \phi_0\sin (k_Az)$ of a field.
Recently, the cosmic variance has been considered in Ref.\cite{Lyth-CMBa}.
Another solution has been found in Ref.\cite{cont-ph}, in which a
contracting phase before slow-roll inflation plays the role.

One of the mysteries of those solutions would be
the origin of the source perturbation $\Delta \phi$, which is supposed to have
specific direction (in this paper we are specifying the direction $z$) in space. 
The cosmic variance~\cite{Lyth-CMBa} could be a solution, but in this
paper we consider topological defects that may appear because of chaotic
initial conditions in the pre-inflationary Universe, and show that the defect
configurations can indeed explain the asymmetry when they are combined with
the curvaton or other modulation mechanisms.\footnote{See also
Ref.~\cite{recent-work, Kanno:2013ohv, Donoghue:2007ze}.}
Here, the curvaton mechanism~\cite{curvaton-original} uses isocurvature
perturbation of a curvaton field $\sigma$.
Although the curvaton is negligible during inflation, the ratio of 
the curvaton to the total energy density grows after inflation and
finally it generates the curvature perturbation.
On the other hand, other modulation mechanisms consider isocurvature perturbation of
a light field (moduli), which is always a negligible fraction of the
energy density.
For instance, the isocurvature perturbation of a light field can make the
decay rate of the inflaton~\cite{modulateddecay} spatially
inhomogeneous, and it can cause generation of curvature perturbation at
the time of reheating.
In the same way, generation of the curvature perturbation is possible
when energy density changes its scaling at phase transition~\cite{IH-pt}.

Another mystery would be the significant scale-dependence of the
asymmetry parameter $A(k)$. 
If one interprets the dipolar asymmetry in the CMB power spectrum 
as a spatial variation of the amplitude of primordial fluctuations,
one can make predictions not just for the CMB but also for large-scale
structure.
Then there should be a corresponding gradient in the number
density of highly biased objects.
The constraint from the Quasar is first obtained in Ref.~\cite{Quasar}.
Using the high-redshift quasars from the Sloan Digital Sky Survey
($z\ge 2.9$), Hirata found a null result for a gradient in the number
density of highly biased objects, which rules out the simple
curvaton-gradient model.
A tighter constraint has been obtained in a recent
paper~\cite{Samuel-CMB}, in which the hemispherical power asymmetry in
the CMB on small angular scales has been investigated.
In Ref.\cite{Samuel-CMB}, it has been shown that the hemispherical power
asymmetry must satisfy $A<0.0045$ on the $10$ Mpc scale. 
In this paper we will analyze this issue using a
simple $\delta N$ formalism and show how to construct
models in which scale-dependent asymmetry meets the criteria.

As we need a simple method for the estimation, we are going to consider
approximation based on the $\delta N$ formalism.
 In Sec.\ref{sec-deltan} we will review previous analysis of the CMB
 asymmetry in the light of the $\delta N$ formalism.
Sec.\ref{Sec:defect} is devoted to the analysis of the asymmetry caused
by a domain-wall configuration.
First, we will show that a simple scenario of topological inflation
cannot explain the anomaly.
Then we will show why the standard modulated decay scenario, as far as
the initial curvature perturbation is mainly sourced by the mechanism, cannot
explain the scale-dependence of the asymmetry.
The same discussion excludes the
curvaton~\cite{Hemisphere-original}.
Although the domain wall configuration considered in this section is
new, other discussions are basically the same as previous
works~\cite{Hemisphere-original, Quasar}.
Then in Sec.\ref{sec-cate}, we will separate 
sources of the cosmological perturbations.
Our idea to achieve the required scale-dependent asymmetry will be
classified in this section.
Then we will examine several specific models in more details.
In addition, we will show that $g_{NL}\ge 10^4$ of the non-Gaussianity
parameter could be crucial.
The origin of the asymmetry parameter will be
discussed in the light a domain wall; however
our method is quite general and can be applied to many other models of
pre-existing configurations~\cite{TiltedTurner}.

\section{$\delta N$ formalism for the CMB asymmetry}
\label{sec-deltan}
During nearly exponential inflation, the vacuum fluctuation of 
each light scalar field $\phi_i$ 
is converted at the horizon exit to a nearly Gaussian
classical perturbation with a spectrum ${\cal P}_{\delta\phi_i}\simeq(H/2\pi)^2$, where 
the Hubble parameter is $H\equiv \dot a(t)/a(t)$.
The curvature perturbation is
\begin{equation}
\zeta = \delta [ \ln (a(x,t)/a(t_*))] \equiv \delta N, 
\end{equation}
where $t$ is time along a comoving thread of space-time and $a(t)$ is the local
scale factor. 
Taking $t_*$ to be an epoch during inflation after relevant scales leave
the horizon, we assume $N(\phi_1(x,t_*),\phi_2(x,t_*),\cdots,t,t_*)$ so that 
\begin{equation}
\zeta(x,t) = N_i \delta \phi_i(x,t_*)
+ \frac{1}{2} N_{ij} \delta \phi_i(x,t_*)\delta \phi_j(x,t_*) + \cdots
, \end{equation}
where a subscript $i$ denotes $\partial/\partial \phi_i$ evaluated on the 
unperturbed trajectory.

We define the fractional power asymmetry as in Eq.(\ref{eq1}), where
$A\sim 0.072\pm 0.022$ for large angular scales ($l<64$) 
is expected from the recent Planck
data~\cite{PlanckCMBA}.\footnote{See also WMAP data~\cite{WMAPCMBA},
which also shows a similar result.} 
In this paper we will examine the possibility of $A\sim 0.05$.

For a single-field perturbation we can expand
\begin{eqnarray}
\frac{\Delta (\delta N)}{\delta N} &=&
 \frac{N_{\phi\phi}}{N_\phi}\Delta \phi +
\frac{1}{2}\frac{N_{\phi\phi\phi}}{N_\phi}(\Delta
\phi)^2+...\nonumber\\
&=& \frac{6}{5}f_{NL} N_\phi\Delta \phi
+\frac{27}{25}g_{NL}\left(N_\phi \Delta \phi\right)^2+...,
\end{eqnarray}
where $f_{NL}$ and $g_{NL}$ are the non-linearity parameters that
measure the non-Gaussianity of the curvature perturbation.
Here we write
\begin{eqnarray}
\label{simple-mod}
\phi(\vec{x},t)&=&\phi_0(t)+\Delta\phi(z,t)+\delta\phi(\vec{x},t),
\end{eqnarray}
where $\delta \phi$ is the conventional Gaussian perturbation and
$\Delta \phi$ is a shift of the field across the sky in the direction of $z$.
Just for simplicity, we are assuming that $\vec{p}$ is in the direction of $z$.
Here we exclude $|N_\phi\Delta \phi|>1$ when $\Delta \phi$ is within the
horizon, since it ruins the perturbative expansion.
Note that in Eq.(\ref{simple-mod}), there are two sources for
perturbation;
 one is the conventional Gaussian perturbation $\delta \phi$ and the other is
the shift $\Delta \phi$.
\subsection{How to introduce scale dependence in the spectrum}
\subsubsection{Curvaton and other modulation mechanisms}
\label{app-running}
In this paper, the scalar field $\varphi$ denotes light fields other than the inflaton.
We are replacing $\varphi\rightarrow\sigma$ for the
curvaton and $\varphi \rightarrow \chi$ for other modulation mechanisms, if these
specifications are possible.

In order to argue the scale-dependence, one must define the
specific time when quantities are evaluated.
Mixings between different definitions will be the source of serious confusions.

The curvaton model uses $\delta \sigma/\sigma$ to define ``component perturbation''
at the beginning of the sinusoidal oscillation:
\begin{equation}
\zeta_\sigma \equiv -H\frac{\delta \rho_\sigma}{\dot{\rho}_\sigma}
=\frac{\delta \rho_\sigma}{3\rho_\sigma}
=\frac{2\delta \sigma}{3\sigma},
\end{equation}
where quantities are defined at the beginning of the oscillation.
Here the curvaton potential is assumed to be quadratic ($\rho_\sigma
=\frac{1}{2}m_\sigma^2\sigma^2$).
In the same way, one can define component perturbation of the radiation.
Evolution of component perturbations is examined in
Ref.\cite{Wands:2000dp}.
Evolution of $\delta \sigma/\sigma$ before the oscillation has been
examined in Ref.\cite{Dimopoulos:2003ss}.

In the curvaton scenario, one usually puts several
assumptions in advance, which makes 
$\delta \sigma/\sigma$ or $\zeta_\sigma$ evolves like constant until the
``event'' (creation of 
the curvature perturbation) takes place. 
In that case, one can calculate the curvaton perturbation using 
\begin{eqnarray}
\label{event-ho}
\zeta_\sigma|_\mathrm{decay}=\zeta_\sigma|_\mathrm{osc}=
\left.\frac{2\delta \sigma}{3\sigma}\right|_\mathrm{osc} &\simeq&
\left.\frac{2\delta \sigma}{3\sigma}\right|_*,
\end{eqnarray}
 where ``$*$'' denotes the time when the Gaussian perturbation of the
corresponding scale ($k=aH$) exits the horizon, and ``decay''  and
``osc'' denote the time of the curvaton decay and the beginning of the
curvaton oscillation, respectively.
Therefore, the curvature perturbation created by the curvaton can be
written as 
\begin{equation}
\zeta\simeq \left[r \zeta_\sigma\right]_\mathrm{decay} 
=[r]_\mathrm{decay}\left[\frac{2\delta \sigma}{3\sigma}\right]_*,
\end{equation}
where $r\equiv 3\rho_\sigma/(3\rho_\sigma+4\rho_r)$.
From the last equation, one can see that the spectrum is strongly scale
dependent when $\sigma_*$ varies significantly during inflation while ${\cal
P}_{\delta \sigma *}\simeq (H_I/2\pi)^2$ is slowly varying.
Obviously, one can see no scale dependence in $r|_\mathrm{decay}$.
On the other hand, if one defines the curvature perturbation using
$\delta \sigma/\sigma|_\mathrm{osc}$, one immediately finds (by
definition)
that $\sigma|_\mathrm{osc}$ cannot be a scale-dependent parameter.
Instead, $\delta \sigma|_\mathrm{osc}$ may have a scale-dependence.
Confusions may arise if one mixes those different definitions.

Similar scale dependence may arise in other modulation mechanisms.
For instance, modulated reheating gives $\zeta\sim \delta
\Gamma/\Gamma\sim \delta \chi/\chi$, where $\Gamma(\chi)$ is the decay
rate that depends on a modulus field $\chi$.
For modulated reheating, ``decay'' dos not mean the decay of the
modulus field but the decay of the inflaton oscillation.

\subsubsection{Scale dependent $\Delta \varphi$ ?}
\label{app-run}

Here we define $\varphi_*\equiv \varphi_0|_*+\Delta \varphi_*$ for a
light scalar field, so that $z$-dependence appears in $\Delta \varphi$.
From the above discussion one will understand that $\Delta \varphi$
could be time dependent so that $\Delta \varphi$ is smaller (if
$V(\varphi)$ is concave) when smaller scales exit horizon.

%

However, if $\varphi$ is the primary source of the initial curvature
perturbation, one will immediately find that the spectral index of the
initial curvature perturbation can put
severe bound on the evolution of $\varphi$.
Since the same is true for $\Delta\varphi$, the scale dependence of
$\Delta \varphi$ must be mild and therefore the scenario of the strongly
scale-dependent $A$ could be 
excluded by using the spectral index. 
The situation becomes better if $\Delta\varphi$ is separated from the
primary source of the initial curvature perturbation. 
Such models will be considered in Sec.\ref{sec-cate}.


\subsection{Example1 : Problem in single-field inflation with a standard kinetic term}
\label{sec-single-our}
Let us apply the $\delta N$ formalism to the simplest scenario.
Here we write for the inflaton field:
\begin{eqnarray}
\phi(\vec{x},t)&=&\phi_0(t)+\Delta\phi(z,t)+\delta\phi(\vec{x},t),
\end{eqnarray}
where $\delta \phi$ is the Gaussian perturbation whose spectrum is ${\cal
P}^{1/2}_{\delta \phi}=H/2\pi$, while $\Delta \phi$ is a shift of $\phi$
across the sky in the direction of $z$.
When one calculates the curvature perturbation using $\delta \phi$, one
has to consider a local mean value $\bar{\phi}(z)\equiv \phi_0+\Delta
\phi(z)$.
Then the $\delta N$ formalism gives,
\begin{eqnarray}
\delta N &=&  N_\phi \delta \phi+\frac{1}{2}N_{\phi\phi}\delta \phi\delta \phi+...
\nonumber\\
&\simeq&\frac{1}{\sqrt{2\epsilon_H}}\frac{\delta \phi}{M_p},
\end{eqnarray}
where $M_p$ is the reduced Planck mass.
Here the slow-roll parameter is $\epsilon(z)_H \equiv-\frac{\dot{H}}{H^2}\simeq
\frac{1}{2}M_p^2\left(\frac{V'}{V}\right)^2$, where $\epsilon_H$ is a
function of $\bar{\phi}(z)$. 
 $M_p=2.436\times 10^8$GeV is the reduced Planck mass and $H\equiv
 \frac{V(\phi)}{3M_p^2}$ 
 is the Hubble parameter during inflation.
The shift $\Delta(\delta N)$ across the sky
is evaluated as
\begin{eqnarray}
\Delta(\delta N)&\simeq&\left(\delta N\right)_\phi\Delta \phi\nonumber\\
&\simeq& N_{\phi\phi}\delta \phi \Delta \phi,
\end{eqnarray}
where the scale dependence of $\delta \phi$ has been neglected. (See also
footnote~\ref{foot-cons}.) 
Then one can easily find
\begin{equation}
A\simeq \frac{N_{\phi\phi}}{N_\phi}|\Delta \phi|.
\end{equation}
Considering the non-Gaussianity parameter
\begin{equation}
f_{NL}\equiv \frac{5}{6}\frac{N_{\phi\phi}}{(N_\phi)^2},
\end{equation}
one will find
\begin{equation}
\label{eq-fnl-plain}
f_{NL}\simeq \frac{5}{6}\frac{A}{N_\phi |\Delta \phi|}.
\end{equation}
For the single-field inflation scenario, a simple observation gives\footnote{\label{foot-cons}
Exact calculation gives $A\propto n_s-1$, where $n_s$ is the
spectral index.
Namely, if one considers $\frac{d}{d\phi}{\cal P}_{\delta
\phi} \ne 0$ and $\epsilon_H\ne 0$, one will find
additional terms that lead to $A\propto n_s-1$.}
\begin{equation}
\label{eq-above}
A\sim 1.2\times f_{NL}N_\phi|\Delta \phi|.
\end{equation}
The above result suggests $\Delta N\equiv N_\phi|\Delta \phi|\sim
(0.05/1.2f_{NL}) \sim 1$ if $|f_{NL}|\sim |n_s-1|$.
Therefore, $\Delta N\sim 1$ may ruin the perturbative expansion.

The above condition may be marginal,
however we can see that the condition becomes more stringent when
 higher terms are considered.
To see the constraints from the higher terms,
let us examine the quadrupole and the octupole of the
perturbation.
First introduce a function $F(k_Az)$, which gives
$F(k_A z_d)\equiv\Delta \phi_d$ on the decoupling scale ($z=z_d$).
For that function, we have the expansion in powers of
$(k_A z)$.
\begin{equation}
\label{eq-expandf}
F(k_A z)=F'(0)(k_A z)
+\frac{1}{2!}F''(k_A z)^2+...,
\end{equation}
where the prime is for the derivative with respect to $(k_A z)$.
In addition to the above expansion, one can expand the gravitational-potential
($\Phi=-\frac{3}{5}\zeta$) in powers of $F$.
Here, we are temporarily considering $\Phi$ instead of $\zeta$, so that the reader can
easily compare the result with the original calculation in
Ref.\cite{Hemisphere-original}.
We first expand $\Delta N_d$ as 
\begin{eqnarray}
\label{eq-expanddeltan}
\Delta N_d&\equiv&N_\phi(\Delta \phi_d)+ 
\frac{1}{2!}N_{\phi\phi}(\Delta \phi_d)^2+...\nonumber\\
&=&N_\phi(\Delta \phi_d)+ 
\frac{6}{10}f_{NL}[N_\phi(\Delta \phi_d)]^2+...
\end{eqnarray}
For a single-field inflation model, in which the non-Gaussianity is
negligible, one can neglect terms proportional 
to $N_{\phi\phi}$ and $N_{\phi\phi\phi}$.\footnote{These terms are not
negligible in the
 curvaton~\cite{curvaton-original, Infcurv0, PBHInfcurv, Infcurv-NonG}
 and other modulation mechanisms~\cite{modulateddecay, modulatedinflation, EKM-KLM,
 Modulated-curvaton, IH-P}. }

For comparison, we are going to introduce a specific function $F(k_A z)
\sim \hat{\phi}\sin(k_A 
z+\omega_0)$, which has been used in Ref.\cite{Hemisphere-original}.
This function will be replaced when we consider a topological defect.
Here $\hat{\phi}$, $k_A$ and $\omega_0$ are constants, which have the corresponding
dimensions.
Using Eq.(\ref{eq-expandf}), one can expand 
\begin{equation}
\label{eq-expand-deltaphi}
\Delta \phi_d= (k_A z_d) \hat{\phi}\cos\omega_0
+\frac{1}{2}(k_A z_d)^2 \hat{\phi}\sin\omega_0+...
\end{equation}
Introducing $\Phi_{\Delta \phi_d}\equiv -\frac{3}{5}\Delta N_d$ and
using Eq.(\ref{eq-expanddeltan}) and (\ref{eq-expand-deltaphi}), one can
expand $\Phi_{\Delta \phi_d}$.
For the first order, we define $\Phi_A$ as 
\begin{equation}
 \Phi_{\Delta \phi_d}|_{\cal D} \equiv (k_A z_d)\Phi_A  \cos\omega_0,
\end{equation}
where the subscript ${\cal D}$ denotes the perturbation proportional to
$k_A z$.
The terms contributing to
the CMB quadrupole and octupole are~\cite{Hemisphere-original}:  
\begin{eqnarray}
\label{qua-oct-cond}
\Phi_{\Delta \phi_d}|_{\cal Q} 
\equiv \frac{(k_A z_d)^2}{2}|\Phi_A \sin\omega_0|&\le&
2.9{\cal Q}\\ 
\Phi_{\Delta \phi_d}|_{\cal O} \equiv \frac{(k_A z_d)^3}{6}|\Phi_A
\cos\omega_0|&\le& 5.3{\cal O},
\end{eqnarray}
where the upper bounds are ${\cal Q}\le 1.8\times 10^{-5}$ and ${\cal O}\le 2.7\times
10^{-5}$ for the quadrupole and the octupole, respectively.
In Ref.\cite{Hemisphere-original}, $\omega_0=0$ has been considered and
thus the quadrupole vanishes.

In the $k_A\rightarrow 0$ limit with fixed $k_A \Phi_A$, one
will find a negligible bound from the quadrupole and the octupole.
In that limit the size of the configuration becomes much larger than the horizon
size and what we are observing in the sky is a local part of the configuration.
Therefore, the configuration is approximately a linear function of $z$.

However, if one introduces the condition ``$\Phi_A\le 1$
everywhere'', it gives $\Phi_{\Delta \phi_d}/(k_A z_d)\le 1$
\footnote{\label{foot-sing}This condition is obviously different from the 
condition ``$|N_\phi\Delta \phi|<1$ within the Horizon''.
The difference will be crucial when we consider a defect that is
expanded during inflation.}.
Then for $\Phi_{\Delta \phi_d}$, we have
\begin{equation}
\label{eq-everywhere}
(k_A z_d)\ge \Phi_{\Delta \phi_d}.
\end{equation}
Therefore, for a fixed $\Phi_{\Delta \phi_d}$ (because we need to
explain $A\sim 0.05$), $k_A$ is bounded from below
and finally we have 
\begin{eqnarray}
|\Phi_{\Delta \phi_d}|^3&\le& 32{\cal O}.
\end{eqnarray}
Here, the quadrupole is neglected assuming $\omega_0=0$.
Using Eq.(\ref{eq-above}), it has been concluded in
Ref.\cite{Hemisphere-original} that a
single-field inflation model will not produce $A\sim 0.05$.

The situation will be changed when $F(k_A z)$ is replaced by a domain
wall configuration and the condition ``$\Phi_A\le 1$
everywhere'' is replaced by ``$\Phi_A\le 1$ within the horizon''.
Let us see more details in the next section.

\section{Topological defects expanded during inflation}
\label{Sec:defect}
In the previous section we have introduced  a planar wave perturbation 
for $\Delta \phi(z)$.
However, it is not quite obvious why such non-spherical perturbation 
has been produced in the inflationary Universe.
In this section we will focus on the source of $\Delta \phi(x)$, paying
attention to chaotic initial conditions in the pre-inflationary
epoch.
The amplitude of the configuration can be as large as the Planck scale.
We are choosing two of the simplest models and will show explicitly how
the defect configurations can affect the asymmetry of the cosmological
perturbations.
More successful (but rather complicated) models will be examined in the
next section so that the model can explain significant scale dependence
of the asymmetry.

\subsection{Inflating Defects}
\label{app-defect}
The idea of topological inflation is very old. 
One can find an excellent review of the cosmological defects in Ref.~\cite{Vilenkin-book}. 

\begin{figure}[ht]
\centering
\includegraphics[width=1.0\columnwidth]{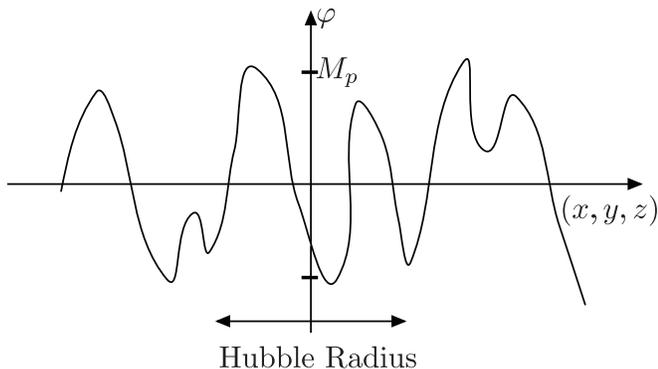}
 \caption{The Universe is inhomogeneous because of the chaotic
 initial condition.} 
\label{fig-defect1}
\end{figure}

To understand the situation, consider a domain wall model for which the 
symmetry is broken by a real field $\phi$ and it develops a vacuum
expectation value $\phi=\pm \hat{\phi}$ at a distance from the core ($\phi=0$).
Just for simplicity, we assume $V(\hat{\phi})=0$ and $V(0)\equiv
V_0>0$.
Before the primordial inflation, we are considering a chaotic initial
condition, which is schematically shown in Fig.\ref{fig-defect1}.

\begin{figure}[ht]
\centering
\includegraphics[width=1.0\columnwidth]{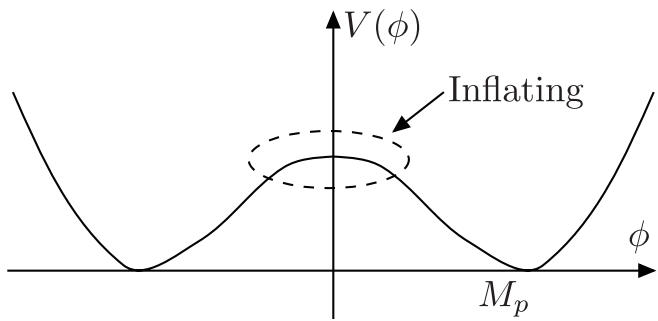}
 \caption{Topological inflation expects a broad core.
What is required for inflation is $\hat{\phi}\ge M_p$.}  
\label{fig-doublewell}
\end{figure}

In flat spacetime the width of the domain wall ($\delta_w$) is
determined by the balance of the gradient and the potential energies
($\delta_w\sim \hat{\phi} /\sqrt{V_0}$).
Since the horizon radius of the Universe when the false vacuum density $V_0$
is dominating is given by $H_0^{-1}\sim M_p/\sqrt{V_0}$, one expects
$\hat{\phi} \ge M_p$ for a trivial (e.g, $\sim -m^2\phi^2+\lambda
\phi^4$) potential.
See also Fig.\ref{fig-doublewell}.
Numerical studies for quartic potentials suggest $\hat{\phi}\simeq
0.33\sqrt{8\pi}M_p$~\cite{Vilenkin-book}.
The Universe first emerged would have a chaotic initial
condition $\dot{\phi}^2\sim (\partial_i\phi)^2\sim M_p^4$, which leads
to a highly inhomogeneous Universe that have a random distribution of
$\phi$ within the Hubble horizon.
At this moment the width of the defects is very narrow since the chaotic
initial condition is (by definition) not determined by the balance
between the gradient and the energies of the potentials.
After a while, the width is determined by the balance of
gradient and potential energies and the topological inflation takes
place.

Here, it must be noted that the curvature perturbation becomes
singular at $\phi=0$, where the
slow-roll parameter vanishes ($\epsilon_H \simeq 0$).
In other words, topological inflation never ends deep in the core.
Slightly away from $\phi=0$, one can find a suitable
region that allows conventional inflation with a safe end accompanied by
the oscillation and reheating.
Obviously, the inflating defect of the topological inflation scenario
leads to a singular perturbation.
However, in reality the singularity is not a problem
since $\phi=0$ is far away from the horizon.
See also footnote~\ref{foot-sing}.

The situation is almost the same for the
curvaton and other modulation mechanisms, but a few
trivial differences may appear.
First, the potential gives $V=V(\varphi)+V_I$ during inflation, where
$V_I\gg V(\varphi)$ is
the energy density of the inflaton field.
The width of the defect can become as large as the horizon radius during
inflation, if the effective mass of $\varphi$ is smaller than the Hubble
parameter ($m_\varphi^2 \le H_I^2$).
{\bf Unlike topological inflation, the Hubble radius is
determined by $V_I$, which is independent of $V(\varphi)$.}
Then the configuration is expanded during the inflationary expansion.
Second, although in this paper we have been considering a specific defect
configuration (domain wall), the gradient of the field ($\Delta
\varphi\ne 0$) can simply be created by the chaotic initial condition,
{\bf even if $\Delta \varphi$ is not related to a topological configuration.}
The gradient $\Delta \varphi$ can be generated on a flat
potential $V(\varphi)\simeq 0$, even if it is not related to a topological defect.
In that case the form of the configuration is determined by the local
shape of the potential $V(\varphi)$.
We have been using a domain wall configuration, as we are expecting that
$F(k_Az)$ is more or less akin to the local configuration of a domain wall.
The initial chaotic condition may include $\varphi=0$,
where conventional perturbation will be singular.
The singularity is not a problem, if it is always far away from our
Universe.

We hope there will be no confusion between defects of
$\phi$(inflaton) and $\varphi$(curvaton/moduli).

\subsection{Example2 : Problem in topological inflation}
The accelerated expansion of the Universe in standard inflationary
scenarios is driven by  the energy of the false vacuum.
Since the topological defects have false vacuum in their cores, one can
expect inflating cores when a defect have (or evolves to have) 
a broad core~\cite{Vilenkin-book, Q-ballinflation}.\footnote{Our idea of
the topological defect is a particular case of a general super
horizon-scale perturbation. See also a preceding work~\cite{TiltedTurner}.
Our method is quite general.}
This gives the basic idea of the inflating defect, which can explain the
initial condition of the conventional inflation model.
All one needs to start inflation is a false vacuum region that is
greater than the horizon.

Assume that our Universe is placed on a primordial domain wall $\phi_w(z) =
\hat{\phi} [\tanh (k_A z+\omega_0)]$.
Expanding the configuration in powers of $(k_Az)$, we find
\begin{eqnarray}
\frac{\phi_w}{\hat{\phi}} &\simeq& \tanh(\omega_0)\nonumber\\
&&+(k_Az) \mathrm{sech}^2(\omega_0)\nonumber\\
&&-(k_Az)^2 \tanh(\omega_0)
 \mathrm{sech}^2(\omega_0)\nonumber\\
&&+\frac{(k_Az)^3}{3} \left[\mathrm{cosh(2\omega_0)-2}\right]
\mathrm{sech}^4(\omega_0)+...,
\end{eqnarray}
where we are going to define $\phi_{CMB}\equiv\hat{\phi}\tanh(\omega_0)$.
The source of the CMB asymmetry is $\Delta \phi \simeq
(k_A z_d)\hat{\phi}\, \mathrm{sech}^2(\omega_0)$.

For hilltop-type inflation~\cite{Hilltop-inflation}, we have
$\hat{\phi}\simeq M_p$ and 
\begin{equation}
\phi_{CMB}\simeq \hat{\phi}e^{-N|\eta|},
\end{equation}
where $\eta\equiv M_p^2 V_{\phi\phi}/V$.
We thus find for $N\sim 60$:
\begin{equation}
\tanh(\omega_0)\simeq e^{-60|\eta|}.
\end{equation}
In the same way as in the previous section, we can calculate the
coefficients of the higher terms to find the quadrupole and the octupole.
Unfortunately however, since the single-field inflation with the
standard kinetic term always 
predicts $f_{NL}\ll 1$, and also the exact calculation of the asymmetry
parameter gives $A\propto n_s-1$, which is mandatory, it is impossible
to find $A\sim 0.05$ from the perturbative expansion.

Although the above result is disappointing, the idea of the
inflating defect is interesting.
Note also that primordial configuration may appear for {\bf every light field}
at the same time, and they could be expanded simultaneously during
inflation, even though the fields themselves do not cause inflationary
expansion.  
Just for an instance, consider a hilltop
curvaton~\cite{Hilltop-curvaton, Topologicalcurvaton},
in which  the curvaton mass is temporarily negative (i.e, the 
curvaton potential is initially a convex) but it finally becomes 
concave before the decay.
If we introduced the chaotic initial condition to the model,
the initial configuration could be a domain wall whose shape
is determined by the potential at that moment.
See also some recent works in Ref.~\cite{refereitem}, in which
authors considered the evolution of both the curvaton
and the modulus field in modulated reheating scenarios and explored the
effects on the power spectrum and $f_{NL}$.

\subsection{Example 3 : Modulated reheating for a domain wall configuration}
In this section we consider a modulus field ($\varphi$), which has both $\Delta
\varphi$ and $\delta \varphi$.
We assume that $\delta \varphi$ is primary source of the initial curvature
perturbation.
In contrast to the topological inflation model considered above,
$\varphi$ is not supposed to be the inflaton field. 
The curvaton mechanism, which has been explored in other
papers~\cite{Hemisphere-original, Lyth-CMBa}, will give a similar result,
although the curvaton is not mentioned explicitly in this section.

The curvaton (except for the inflating curvaton) and other modulation
mechanisms (e.g, modulated reheating) do not expect $|f_{NL}|\ll 1$ even if the decaying 
matter component dominates the Universe~\cite{modulateddecay, EKM-KLM}. 
In this section we consider a modulated decay scenario that gives
$|f_{NL}|\sim 5$~\cite{EKM-KLM}. 
For a moduli field $\varphi$, we thus find from Eq.(\ref{eq-above}):
\begin{equation}
\label{eq-mod-1}
A\sim 6 N_\varphi|\Delta \varphi|.
\end{equation}
Here, we assume that the origin of $\Delta \varphi$ is the primordial
domain wall that exists
prior to the inflationary expansion.
One can define the mean value (center
of the Gaussian perturbation) of the
moduli as $\bar{\varphi}\equiv 
\varphi_0+\hat{\varphi}\tanh(\omega_0)+\Delta \varphi$.
Here the domain-wall configuration $\varphi_w\equiv
\hat{\varphi}\tanh(k_A z+\omega_0)$ is centered at $\varphi_0$ and is 
expanded for $k_A z\ll 1 $ as 
\begin{eqnarray}
\frac{\varphi_w}{\hat{\varphi}} &\simeq& \tanh(\omega_0)\nonumber\\
&&+(kz) \mathrm{sech}^2(\omega_0)\nonumber\\
&&-(kz)^2 \tanh(\omega_0)
 \mathrm{sech}^2(\omega_0)\nonumber\\
&&+\frac{(kz)^3}{3} \left[\mathrm{cosh(2\omega_0)-2}\right]
\mathrm{sech}^4(\omega_0)+...,
\end{eqnarray}
Therefore, $\Delta \varphi$ is
\begin{eqnarray}
\frac{\Delta \varphi}{\hat{\varphi}}&\simeq&
(kz) \mathrm{sech}^2(\omega_0)
-(kz)^2 \tanh(\omega_0)
 \mathrm{sech}^2(\omega_0)...\nonumber\\
\end{eqnarray}
The expansion by $\Delta \varphi$ is
\begin{equation}
\Delta N=N_\varphi\Delta \varphi + \frac{1}{2}N_{\varphi\varphi}(\Delta
 \varphi)^2
+...
\end{equation}
Therefore, the dipole (not observable), the quadrupole and the octupole are
\begin{eqnarray}
\label{eq-d}
\frac{\Delta N|_{\cal D}}{k_Az_d}&\equiv& N_\varphi
 \hat{\varphi}\,\mathrm{sech}^2(\omega_0)\equiv \Delta N_A\\
\label{eq-q}
\frac{\Delta N|_{\cal Q}}{(k_Az_d)^2}&\equiv&
\frac{1}{2}N_{\varphi\varphi}\hat{\varphi}^2\,\mathrm{sech}^4(\omega_0) 
-N_\varphi \hat{\varphi}\tanh(\omega_0) \mathrm{sech}^2(\omega_0)\nonumber\\
&=&\frac{6}{10}f_{NL}(\Delta
	  N_A)^2-\Delta N_A\tanh(\omega_0)\\
\frac{\Delta N|_{\cal O}}{(k_Az_d)^3}&\equiv&
\frac{1}{6}N_{\varphi\varphi\varphi}\hat{\varphi}^3\,\mathrm{sech}^6(\omega_0) 
\nonumber\\
&&-N_{\varphi\varphi}\hat{\varphi}^2\,\mathrm{sech}^4(\omega_0) 
\tanh(\omega_0)\nonumber\\
&&+\frac{1}{3}N_\varphi \hat{\varphi}\left[\mathrm{cosh(2\omega_0)-2}\right]
\mathrm{sech}^4(\omega_0)\nonumber\\
&=&\frac{9}{25}g_{NL}(\Delta
	  N_A)^3-\frac{6}{5}f_{NL}\tanh(\omega_0)(\Delta N_A)^2\nonumber\\
&&+\frac{\Delta N_A}{3} \left[\mathrm{cosh(2\omega_0)-2}\right]
\mathrm{sech}^2(\omega_0).
\end{eqnarray}
From Eq.(\ref{eq-above}) and Eq.(\ref{eq-d}), we find
\begin{equation}
A\sim 1.2 f_{NL} \Delta N_A (k_A z_d).
\end{equation}
From Eq.(\ref{eq-q}), we find 
\begin{equation}
\label{eqcond2}
f_{NL}(\Delta N_A)^2(k_Az_d)^2<8{\cal Q}.
\end{equation}
Therefore, the asymmetry is bounded from above as
\begin{equation}
A < 1.2 f_{NL}\sqrt{\frac{8Q}{f_{NL}}}\sim 0.014 \sqrt{f_{NL}}.
\end{equation}

Since $g_{NL}> 10^4$ is not excluded in the modulated decay scenario 
and is giving an interesting observational possibility,
an important issue is to consider a more stringent constraint that may appear
from the octupole perturbation 
\begin{eqnarray}
\frac{\Delta N|_{\cal O}}{(k_Az_d)^3}
&\simeq&\frac{9}{25}g_{NL}(\Delta N_A)^3< 8.8 {\cal O},
\end{eqnarray}
which will be significant if $|g_{NL}|^2\gtrsim 10^5 |f_{NL}|^3$.
We thus find that an observation of $g_{NL}>10^4$ could be crucial for
the models of the CMB asymmetry.  
This point has not been considered in previous works.
 
One might expect that the significant scale dependence of the
asymmetry parameter $A$ could be explained by the scale-dependent $f_{NL}$.
However, usual scenario of the curvaton and other modulation mechanisms
do not expect such significant scale dependence.
Moreover, since $f_{NL}$ has the minimum value ($f_{NL}\gtrsim 1$) associated with nonlinear
effects in any model~\cite{EKM-KLM}, the variation stops inevitably there. 
On the other hand, $f_{NL}$ has the upper bound $|f_{NL}|\le 10$ on large
scale, which can contradict with the required scale
dependence.\footnote{The Planck limit assumes that $|f_{NL}|$ is constant
on all CMB scales. Although we did not consider the possibility in this
paper, $|f_{NL}|$ might have a strong scale dependence and $|f_{NL}|$ on the scale of the asymmetry could be far larger than
$|f_{NL}|\sim 10$.}
Namely, to solve the scale dependence of the asymmetry using
scale-dependent $f_{NL}$, one needs 
$10/|f_{NL}^{(10\mathrm{Mpc})}| \ge 0.05/0.0045 $, which gives a
critical condition
$|f_{NL}^{(10\mathrm{Mpc})}| <1$.
The inflating curvaton could be an exception, in which
$f_{NL} \sim {\cal O}(\epsilon_H, \eta)\ll 1$ is possible.
However, at this moment there is no concrete model that realizes
strongly scale-dependent $f_{NL}$ in the inflating curvaton model.
Therefore, we conclude that all these models cannot explain the asymmetry.

\section{Model buildings}
\label{sec-cate}

Let us sort out cosmological models paying
attention to the origin of the asymmetry and its scale dependence.
Prescriptively, ``multi-field models'' include the curvaton and other
modulation mechanisms since these models require at least two fields to
achieve both the 
inflationary expansion and the creation of the curvature
perturbations.

As we have stated in the previous section, ``single-field models of
inflation'' are already excluded if the source of the asymmetry is $\Delta \phi(z)\ne 0$.
The curvaton and other modulation mechanisms, which are prescriptively
multi-field but one field (the curvaton $\sigma$ or a modulus field
$\chi$) is the primary source of the initial curvature
perturbations, can be excluded by the scale dependence.

Therefore, in this section we consider separation of $\delta N$.
The separation of the curvature perturbation ($\delta N=\delta
N_1+\delta N_2$) is a very old idea, which has been considered in
a variety of multi-field models of inflationary cosmology. 
Ref.\cite{Hemisphere-original} considers an application to explain
the asymmetry, and Ref.\cite{running-inf-asym} calculates the
scale-dependence of the parameter $A$, when the slow-roll parameter is
scale-dependent.
The crucial difference between previous works and our model will be
explained in Appendix \ref{app-review-EHK}.

In this section we need strong scale dependence for the secondary
component $\delta N_2$.
Recently in Ref.\cite{PBHcurvaton}, strongly scale-dependent spectrum of the
curvaton scenario has been used to explain the primordial black hole
(PBH) formation. 
In Ref.\cite{PBHInfcurv}, it has been shown that the inflating
curvaton mechanism can also explain PBH formation, in which some
constraints can be relaxed because of the late-time curvaton inflation.
Other modulation mechanism may also have strong scale dependence.
For instance, a model of the scale-dependent modulations (tachyonic growth
model) has been proposed in Ref.\cite{John-tach}.

Suppose that the potential of the field $\varphi$ is given by
$V(\varphi)\simeq cH^2 \varphi^2$.
The potential is familiar in supergravity inflationary models, in which
F-term generates  $|c|\sim {\cal O}(1)$ while D-term may generate $|c|\ll 1$.
It is possible to generate $V(\varphi)\simeq cH^2 \varphi^2$ from 
$F(\varphi)R$ in the Lagrangian, where $F(\varphi)$ is a function of $\varphi$ and $R$ is
the Ricci scalar.
In that way, $c$ is determined by the component of the Universe.
One will find the equation
\begin{equation}
\ddot{\varphi}+3H\dot{\varphi}+cH^2\varphi=0,
\end{equation}
whose solution is~\cite{Dimopoulos:2003ss}
\begin{equation}
\varphi\propto e^{-\alpha Ht}\propto k^{-\alpha},
\end{equation}
where $\alpha\equiv \frac{3}{2}-\sqrt{\frac{9}{4}-c}$. 
If $\delta \varphi$ is primary source of the initial curvature
perturbation, contribution to the spectral index is
$n_s-1=-2\epsilon_H+2\alpha\simeq 2\epsilon_H+2\eta_\varphi$, where
$\eta_\varphi\equiv M_p^2 V_{\varphi\varphi}/V$.

\subsection{Multi-Field models of separable perturbations
 ($N_{\phi\varphi}\simeq 0$)}

First, consider a separable spectrum of $\phi$ and $\varphi$
that gives
\begin{equation}
\delta N= \delta N_1 + \delta N_2,
\end{equation}
where we defined 
\begin{eqnarray}
\delta N_1&=& N_\phi \delta \phi\\
\delta N_2 &=& N_\varphi \delta \varphi.
\end{eqnarray}
To be separable up to the second order, we need
 $N_{\phi\varphi}\simeq0$.

We also assume a Gaussian perturbation and a shift given by
\begin{eqnarray}
\phi(\vec{x},t)&=&\phi_0(t)+\delta\phi(\vec{x},t),\nonumber\\
\varphi(\vec{x},t)&=&\varphi_0(t)+\Delta\varphi(\vec{x},t)+\delta\varphi(\vec{x},t),
\end{eqnarray}
where $\delta \phi$ and $\delta \varphi$ are the Gaussian perturbation,
which is expected to have the spectrum ${\cal 
P}_{\delta \phi}^{1/2}={\cal
P}^{1/2}_{\delta \varphi}=H_I/2\pi$ at the horizon exit.
As before, we introduce a function $F(k_A z)$ that gives 
$F(k_A z_d)\equiv \Delta \varphi_d$ on the decoupling scale.
Then, one can define the asymmetry as
\begin{eqnarray}
A &\equiv& \frac{\Delta(\delta N_2)}{\delta N}\nonumber\\
&=& \frac{N_{\varphi\varphi}{\cal P}^{1/2}_{\delta \varphi} |\Delta \varphi|}
{N_\phi {\cal P}^{1/2}_{\delta \phi}+
N_\varphi {\cal P}^{1/2}_{\delta \varphi}}.
\end{eqnarray}

\begin{figure}[ht]
\centering
\includegraphics[width=1.0\columnwidth]{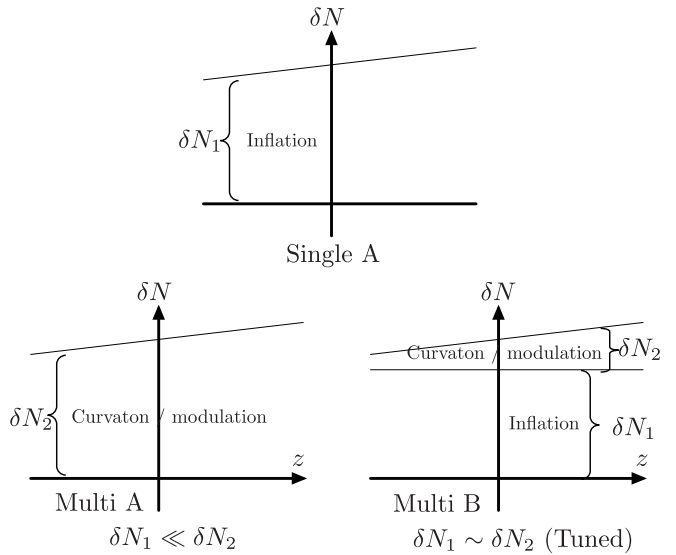}
 \caption{The left-hand side picture shows a model in which the Gaussian
 perturbation of a field is the primary source of the initial curvature
 perturbations and at the same time defect configuration of the field
 explains the asymmetry of the CMB.
Although somewhat confusing, the model is usually dubbed
 ``multi-field'', since one needs two fields 
to explain both the inflationary expansion and the curvature
 perturbation.
The right-hand side picture shows a model in which the initial curvature
 perturbation is generated by a field but the source of the asymmetry is 
another field. 
} 
\label{fig:two-models}
\end{figure}

We are going to separate ``multi-field models of separable
perturbations'' into two categories. 
From Fig.\ref{fig:two-models}, ``Multi A'' includes the conventional curvaton and
modulations, in which a field $\varphi$ is
responsible for ``both'' the initial curvature perturbations and the
asymmetry.
In that case we find 
\begin{eqnarray}
A &\simeq& \frac{N_{\varphi\varphi} |\Delta \varphi|}
{N_\varphi}.
\end{eqnarray}

The other (``Multi B'') considers one field ($\phi$) as primary source
of the initial curvature perturbations and the other field ($\varphi$)
as the source of the asymmetry.
In that case we are expecting $\delta N_2 < \delta N_1$ but not
much smaller.
Since the asymmetry is expected to be $A\sim 0.05$, one can estimate
$\delta N_2/\delta N_1\ge A\sim 0.05$.
{\bf Later we will show that $\delta N_2/\delta N_1 > 0.05$ 
could be a crucial condition.}
For this model, the asymmetry is defined by
\begin{eqnarray}
A &\simeq& \frac{N_{\varphi\varphi}{\cal P}_{\delta\varphi}^{1/2} |\Delta \varphi|}
{N_\phi {\cal P}^{1/2}_{\delta \phi}}\nonumber\\
&=& \frac{N_{\varphi\varphi} |\Delta \varphi|}
{N_\phi} \left(\frac{{\cal P}_{\delta \varphi}}{{\cal P}_{\delta \phi}}\right)^{1/2},
\end{eqnarray}
where the quantities are defined at the moment when the curvature perturbations
are generated. (i.e, ${\cal P}_{\delta \phi}$ is defined at the
horizon exit but ${\cal P}_{\delta \varphi}$ is defined when $\delta N_2$
is generated.)
Then, following the discussion in Sec. \ref{app-running}, significant
scale-dependence may appear in the ratio  
$ \left(\frac{{\cal P}_{\delta \varphi}}{{\cal P}_{\delta
\phi}}\right)^{1/2}$.
In order to explain the scale dependence in a familiar form,
consider the function
\begin{eqnarray}
\varphi &\equiv& g(\varphi_*),
\end{eqnarray}
where the definition first appeared in Ref.\cite{Lyth-gfc} to explain
the evolution of the curvaton perturbation.
We thus find
\begin{eqnarray}
A &\simeq& \frac{N_{\varphi\varphi} |\Delta \varphi|}
{N_\phi} \left(\frac{1}{g'}\right)^{1/2},
\end{eqnarray}
where $\delta \varphi \simeq g'(\varphi_*)\delta \varphi_*$ and 
${\cal P}_{\delta\phi_*}={\cal P}_{\delta \varphi_*}$ are considered.
The scale-dependent asymmetry is due to the significant
scale dependence of $\delta N_2$.\footnote{Our model (multi-B) does not include the
running inflation scenario.
Thanks to the referee of JCAP, we found that the 
scenario of running inflation has already been thoroughly explored by Erickcek, 
Hirata and Kamionkowski in the appendix A of
Ref.\cite{running-inf-asym}, where the spectrum with a discontinuity has
also been considered.
However, the model assumes $\varphi_*=g(\varphi_*)$ (or
$\eta_\sigma \equiv m_\sigma^2/3H^2\simeq 0$ for the curvaton) and calculated
the asymmetry with $1/g'\equiv1$ (i.e, they considered the trivial evolution
function and put ${\cal P}_{\delta \sigma}={\cal P}_{\delta \sigma_*}$ from
the beginning). The calculation is reviewed in our appendix \ref{app-review-EHK},
where the correspondence between these calculations will be very clear.} 

\subsubsection{Curvaton and other modulation mechanisms (Multi-B)}
\label{sec-multi-B}
Here we consider Multi-B scenario, which is shown on the right-hand side
in Fig.\ref{fig:two-models}.\footnote{Just before we accomplish the second
version of this paper, we found a peper~\cite{John-mod} in which a
similar idea has been examined.}
If the asymmetry of the CMB is created by a field that is NOT primary
source of the initial curvature 
perturbation, the role of the secondary 
field $(\varphi)$ is to create the asymmetric part $\delta N_2\propto
z$.

To begin with, define the non-Gaussianity parameter of the component
perturbation as
\begin{eqnarray}
f_{NL,\varphi}&\equiv&
 \frac{5}{6}\frac{N_{\varphi\varphi}}{(N_\varphi)^2}.
\end{eqnarray}
Then the asymmetry parameter is
\begin{eqnarray}
A &\equiv& \frac{\Delta(\delta N_2)}{\delta N}\nonumber\\
&=& \frac{6}{5}\frac{f_{NL,\varphi} N_\varphi|\Delta \varphi|}
{1+ r_N^{-1} r_{\cal P}^{-1}}\nonumber\\
&\simeq& \frac{6}{5}f_{NL\varphi} N_\varphi|\Delta \varphi|(r_N r_{\cal P}),
\end{eqnarray}
where we defined 
\begin{eqnarray}
r_N^{-1}&\equiv& \frac{N_\phi}{N_\varphi}\\
r_{\cal P}^{-1}&\equiv& \frac{{\cal P}^{1/2}_{\delta \phi}}{{\cal
 P}^{1/2}_{\delta \varphi}}.
\end{eqnarray}
We used the condition $\delta N_1 >\delta N_2$
 that leads to $r_N^{-1} r_{\cal P}^{-1}>1$.

Since the secondary perturbation ($\delta N_2$) is responsible for the
asymmetry, what we need to explain $A=0.05$ is 
\begin{eqnarray}
\label{Comp-AfNL}
f_{NL,\varphi}&\sim& \frac{5}{6}\frac{A}{N_\varphi \Delta \varphi}
r_N^{-1}r_{\cal P}^{-1}\nonumber\\
&\le & \frac{5}{6}\frac{1}{N_\varphi \Delta \varphi},
\end{eqnarray}
where we considered $1>\delta N_2/\delta N_1 >A$.

Then, the non-Gaussianity parameter of the total curvature perturbation is
\begin{eqnarray}
\label{eq-multib-fnl}
f_{NL}&=&\frac{f_{NL,\phi}}{1+r_N^2r_{\cal P}^2} +
 \frac{f_{NL,\varphi}}{1+(r_Nr_{\cal P})^{-2}}\nonumber\\
&\sim & (r_Nr_{\cal P})^2 f_{NL,\varphi}\nonumber\\
&\ge & \frac{5}{6}\frac{A^2}{N_\varphi \Delta \varphi},
\end{eqnarray}
where the second line is obtained assuming $f_{NL,\phi}\equiv \frac{5}{6}\frac{N_{\phi\phi}}{N_\phi^2}\simeq 0$.
Note that in the above equation there is 
significant suppression due to an extra factor of
$A\sim 0.05$.
(Compare Eq.(\ref{eq-multib-fnl}) with Eq.(\ref{eq-fnl-plain}).)
Therefore, the required $f_{NL}$ can be much smaller than the Multi-A
scenario.

We find for the
domain-wall configuration:
\begin{eqnarray}
\frac{\Delta N|_{\cal D}}{k_Az_d}&\equiv& N_\varphi
 \hat{\varphi}\,\mathrm{sech}^2(\omega_0)\equiv \Delta N_A\\
\frac{\Delta N|_{\cal Q}}{(k_Az_d)^2}&\equiv&
\frac{1}{2}N_{\varphi\varphi}\hat{\varphi}^2\,\mathrm{sech}^4(\omega_0) 
-N_\varphi \hat{\varphi}\tanh(\omega_0) \mathrm{sech}^2(\omega_0)\nonumber\\
&=&\frac{6}{10}f_{NL,\varphi}(\Delta
	  N_A)^2-\Delta N_A\tanh(\omega_0)\\
\frac{\Delta N|_{\cal O}}{(k_Az_d)^3}&\equiv&
\frac{1}{6}N_{\varphi\varphi\varphi}\hat{\varphi}^3\,\mathrm{sech}^6(\omega_0) 
\nonumber\\
&&-N_{\varphi\varphi}\hat{\varphi}^2\,\mathrm{sech}^4(\omega_0) 
\tanh(\omega_0)\nonumber\\
&&+\frac{1}{3}N_\varphi \hat{\varphi}\left[\mathrm{cosh(2\omega_0)-2}\right]
\mathrm{sech}^4(\omega_0)\nonumber\\
&=&\frac{9}{25}g_{NL,\varphi}(\Delta
	  N_A)^3-\frac{6}{5}f_{NL}\tanh(\omega_0)(\Delta N_A)^2\nonumber\\
&&+\frac{\Delta N_A}{3} \left[\mathrm{cosh(2\omega_0)-2}\right]
\mathrm{sech}^2(\omega_0).
\end{eqnarray}
Therefore, we find
\begin{equation}
f_{NL,\varphi}(\Delta N_A)^2(k_Az_d)^2<8{\cal Q}.
\end{equation}
Using Eq.(\ref{Comp-AfNL}), we find that the asymmetry is bounded from above as
\begin{equation}
\label{result-b}
A<0.014 \sqrt{f_{NL,\varphi}} (r_Nr_{\cal P}).
\end{equation}
Then $A\sim 0.05$ gives $3.6 <\sqrt{f_{NL,\varphi}}(r_Nr_{\cal P})$.

To explain the scale-dependence of the asymmetry, we consider (again) a variation of 
$r_{\cal P}$.
Here, what we need is $(r_{\cal P}^\mathrm{small}/r_{\cal
P}^\mathrm{CMB})\lesssim 0.1$, which is quite conceivable if the factor is obtained
from the variation caused by a quadratic potential:
$e^{-\eta_\varphi N}\sim 10$ for $N=6.1$ (Quasar) or $N\sim 4$ (CMB
small scale).
Here the slow-roll parameter required for the Quasar is
\begin{equation}
\eta_\varphi\equiv \frac{m_\varphi^2}{3H^2}\sim -0.38.
\end{equation}
For the small-scale CMB, it becomes $\eta_\varphi \sim
-0.58$.\footnote{If $\eta_\varphi>3/2$ then the field will not be
perturbed during inflation~\cite{limit-eta}.}


In this scenario, there will be a possible tension between the spectral
index and the scale-dependence in $\delta N_2$, where the latter is needed to explain the
asymmetry of the small-scale perturbations while
the former is strictly constrained by the Planck observation.
In the above scenario, we can estimate the additional contribution to
the spectral index as (for the Quasar):
\begin{eqnarray}
\Delta (n_s-1)&\sim& \frac{\delta N_2}{\delta N_1}(n_s-1)_{\delta N_2}\nonumber\\
&\ge& 2A \eta_\varphi\sim -0.038,
\end{eqnarray}
where $(n_s-1)_{\delta N_2}$ is the spectral index of the component
$\delta N_2$.
The small-scale CMB constraint~\cite{Samuel-CMB} requires
$n_s  \sim (n_s)_{\delta N_1} +\Delta (n_s)\sim 0.96$
for $\Delta (n_s-1)\sim 2A \eta_\varphi\sim -0.058$, which cannot be
satisfied without fine-tuning.
Therefore, in reality  $\Delta (n_s-1)$ is larger than $2A\eta_\varphi$
 and the issue of the fine-tuning is serious.

\subsection{Multi-field models of mixed perturbations
  ($N_{\phi\varphi}\ne 0$)}
\label{sec-multi-nonsep}
To start with, assume a primary field ($\phi$) and a mechanism
(inflation, the curvaton or other modulation mechanisms) that creates curvature
perturbation $\zeta=\delta N \sim N_\phi\delta \phi$.
Then, one can introduce a modulation sourced by a secondary field $\varphi$ (moduli).
We write
\begin{eqnarray}
\phi(\vec{x},t)&=&\phi_0(t)+\delta\phi(\vec{x},t)\nonumber\\
\varphi(\vec{x},t)&=&\varphi_0(t)+\Delta\varphi(\vec{x},t)+\delta\varphi(\vec{x},t),
\end{eqnarray}
where $\delta \phi$ is the Gaussian perturbation, 
while $\Delta \varphi$ has specific direction $(z)$ and it is 
supposed to be the source of the CMB asymmetry $A\ne 0$. 

A shift $\Delta(\delta N)$ appears because of $\Delta
\varphi$:
\begin{equation}
\Delta(\delta N)\simeq\left(\delta N\right)_\varphi\Delta \varphi
\simeq N_{\phi\varphi}\delta \phi \Delta \varphi.
\end{equation}
Then the asymmetry parameter is
\begin{equation}
\label{A-mixed}
A\simeq \frac{N_{\phi\varphi}}{N_\phi} |\Delta \varphi|.
\end{equation}
Recall that the usual definition of the non-Gaussianity parameter is~\cite{Lyth-gfc}
\begin{equation}
\label{originalfnl}
f_{NL}\equiv \frac{5}{6}\frac{N_a N_b N^{ab}}{(N_c N^c)^2},
\end{equation}
which is obtained assuming $r_{\cal P}=1$.
Therefore we have
\begin{equation}
\label{fnl-mixed}
f_{NL}\simeq \frac{5}{3}\frac{N_\phi N_\varphi N^{\phi\varphi}}
{(N_\phi^2+N_\varphi^2)^2},
\end{equation}
where $N^{\phi\varphi}=N^{\varphi\phi}$.
From Eq.(\ref{A-mixed}) and (\ref{fnl-mixed}), the asymmetry will be
estimated as
\begin{eqnarray}
A&\sim&\frac{N_{\phi\varphi}|\Delta \varphi|}{N_\phi}\nonumber\\
&\sim&\frac{3}{5}f_{NL}\frac{(1+ r_N^2)^2}{
r_N^2}N_\varphi|\Delta \varphi|\nonumber\\
&\sim& \frac{3}{5}\frac{1}{ r_N^2}
 f_{NL}  N_\varphi|\Delta \varphi|.
\end{eqnarray}
The mixed perturbation $N_{\phi\varphi}\ne 0$ may appear when one
considers modulated decay for the curvaton~\cite{Modulated-curvaton,
EKM-KLM}.
Just for an instance, introduce a moduli-dependent mass $m(\phi,\varphi)$ for the decaying
matter component and consider modulated curvaton decay~\cite{EKM-KLM,
Modulated-curvaton}.  
Then one will find $N_\phi$ and
$N_{\phi\varphi}$ in terms of $m(\phi,\varphi)$.
In that case, choosing a point on the landscape of
$m(\phi,\varphi)$, one may find a significant $N_{\phi\varphi}$ that
determines the asymmetry.
A simple modulation of an inflation parameter may lead to the
asymmetry~\cite{modulatedinflation}, which will be considered
below.\footnote{A stringent condition has been found in
Ref.\cite{Kanno:2013ohv}, in which they argued that $N_\varphi \ge
N_{\varphi\phi}\dot{\phi}H^{-1}$ can break a constraint
\cite{PlanckCMBA}.
However, true inequality is $N_\varphi \ge N_{\varphi\phi}\dot{\phi}H^{-1}
+N_{\varphi\varphi}\dot{\varphi}H^{-1}$, which allows cancellation
between terms. Therefore, the mixed perturbation
scenario is not excluded but it may require fine-tuning.}

\subsubsection{Two-field inflation with a non-standard kinetic term ($c_s$-modulation)}
First we review the preceding model considered in
Ref.\cite{c_s-modulation}.
Although the model is slightly complicated compared with our model in
section \ref{sec-our-mized},
the kinetic term is well motivated in the string theory.

In section~\ref{sec-single-our} we showed that ``the single-field inflation model
with the standard kinetic term'' is severely constrained and cannot
explain the asymmetry.
On the other hand, a non-standard kinetic term may change the sound
speed (usually labeled by $c_s$), which causes deviation from the conventional
model of single-field inflation.
The most obvious character of the model appears in the equilateral 
non-Gaussianity parameter $f_{NL}^{eq}$, which can be enhanced 
keeping the spectral index small.
The reason has been clearly explained in Ref.~\cite{DBI-inthesky}, which
shows that the shrinking sound horizon during inflation 
cancels the deviation from a de Sitter background to retain a scale
invariant spectrum.
In that way the enhancement of the non-Gaussianity parameter is possible
without violating the conditions for the spectral index.

Unfortunately, the model with $c_s\ll 1$ has been excluded
by the Planck observation; that is the reason why we did not consider the
model in section~\ref{sec-single-our}.
However, this model may be applied to explain the hemispherical
asymmetry if the value of sound speed varies from one side of the sky to
the other (i.e, when $c_s(k,z)$ is a non-trivial function of both $k$
(scale) and $z$ (place).)
The example considered in Ref.\cite{c_s-modulation} uses the Lagrangian
for the two-field model:
\begin{eqnarray}
 {\cal L} &=& \frac{1}{f(\phi, \chi)}
  (1-\sqrt{1+f\partial_\mu\phi\partial^\mu\phi})\nonumber\\
&& - \frac{1}{2}\partial_\mu\chi\partial^\mu\chi - V(\phi, \chi).
\end{eqnarray}
Here, $\phi$ is the inflaton and $\chi$ is an extra light field
(moduli), which does not contribute to the background evolution.
$\chi$ is introduced just to explain $k$- and $z$-dependent sound speed
$c_s(k,z)$. 
The kinetic term of the inflaton $\phi$ involves a function
$f (\phi, \chi)$.
Since we do not have new result for this model, we will not examine this
model in this paper.
See Ref.\cite{c_s-modulation} for further discussions.

\subsubsection{Modulated inflation with a non-standard kinetic term}
\label{sec-our-mized}
We introduce a moduli field ($\chi$) and consider the non-standard 
kinetic term for the inflaton~\cite{Modulated-nonK}:\footnote{See also
appendix \ref{app-mod}.}
\begin{eqnarray}
{\cal L}_\mathrm{kin}&=&\frac{1}{2}\omega(\chi)
\partial_\mu\phi\partial^\mu \phi.
\end{eqnarray}
Because of $\omega$, the equation of motion becomes
\begin{equation}
\ddot{\phi}+3H\dot{\phi}+\frac{V_\phi}{\omega}+
\frac{\omega_\chi}{\omega}\dot{\phi}{\dot{\chi}}=0,
\end{equation}
where we can assume $\dot{\chi}\simeq 0$.

Let us examine if this simplest version of the non-standard kinetic term
could work to explain the CMB
asymmetry.

The curvature perturbation generated during inflation is
\begin{eqnarray}
\delta N &=&  N_\phi \delta \phi+\frac{1}{2}N_{\phi\phi}\delta \phi\delta \phi+...
\nonumber\\
&\simeq&H\frac{\delta
 \phi}{\dot{\phi}}=\frac{\omega}{\sqrt{2\epsilon_H}}\frac{\delta \phi}{M_p}.
\end{eqnarray}
Since the kinetic term is modulated by $\chi$, one
can expect an inhomogeneous $\dot{\phi}$ in the direction of $z$.
If $\chi$ has a domain-wall configuration ($\Delta \chi(z)$), one can
expect asymmetry caused by the shift.
The asymmetry is
\begin{eqnarray}
A&\simeq&  \frac{N_{\phi\chi}}{N_\phi}|\Delta \chi|\nonumber\\
&=& \frac{\omega_\chi}{\omega}|\Delta \chi|.
\end{eqnarray}
For a specific function, Eq.(\ref{eq-last}) suggests 
\begin{equation}
A\simeq 2\frac{\Delta \chi}{\chi}.
\end{equation}
This is a convincing way of obtaining asymmetry from a very simple
extension of the single-field inflation model.

The scale-dependence can appear from the evolution of
$\chi_*\equiv \chi^{(0)}_*+\Delta \chi_*$, where $\delta \chi$ is
neglected for simplicity.
In this model, scale dependence of the asymmetric perturbation 
$\Delta (\delta N)\sim N_{\phi\chi} \Delta \chi$ can be {\bf
independent} of the dynamics of $\phi$.
This extension can be applied to many kinds of single-field inflationary
models without changing the predictions of the original model, as far as the
model predicts negligible $N_\chi$.
On the otherhand, the predictions of the model are very weak, since the
effective action is quite ambiguous.

\section{Conclusion and discussion}
\label{conclusion}
In this paper, we considered a chaotic initial condition and a domain
wall configuration that may exist prior to the inflationary expansion.
We showed that such configuration can explain the CMB asymmetry.
Remembering that topological inflation may usually include a point where the
curvature perturbation diverges ($\epsilon_H \simeq 0$), we have to
reconsider the conventional assumption ``$\Phi_A\le 1$ everywhere''.
In this paper, the condition has been replaced by ``$N_\phi\Delta \phi\le
1$ (or $N_\varphi\Delta \varphi\le 1$) within the horizon''. 
Although the new condition is looser than the former, we found that
topological inflation cannot explain the asymmetry.
Then we explored the curvaton and other modulation mechanisms using a
primordial defect configuration as the source of $\Delta \varphi(z)$.
The scale-dependence of the asymmetry is examined
in various scenarios.

Our result shows that Multi-A model, in which $\varphi$ is primary
source of the initial curvature perturbation, cannot explain the
asymmetry (and its scale dependence).

On the other hand, Multi-B model, in which $\Delta \varphi$ is separated from
the primary source of the initial curvature perturbation, can explain
both the asymmetry and its scale dependence but requires fine-tuning.
For the spectral index, the reason is very clear.
Since the asymmetric part is entirely due to $\delta N_2$, the fraction $\delta
N_2/\delta N_1$ cannot be smaller than $A\sim 0.05$.
Since the scale dependence of $A$ is due to the significant scale
dependence of $\delta N_2$, it can contribute to the spectral index of
the total curvature perturbation ($\delta N=\delta N_1+\delta N_2$).
If $\delta N_2$ has the spectral index $(n_{s2}-1)\sim {\cal O}(1)$, its
contribution to the total ($n_s-1$) is more than $A\times (n_{s2}-1)\sim
0.05$, which is not acceptable without fine-tuning.
Large $g_{NL}$, which might be observed in future observations, may put a
stringent bound on the octupole perturbation.

We also examined mixed perturbation models.
(In this case we don't have to assume $\delta N_2/\delta N_1>A$.)
$c_s$-modulation has been considered in Ref.~\cite{c_s-modulation}.
In this paper we considered a modulated inflation
model~\cite{modulatedinflation} and showed that the model does not
require fine-tuning in the spectral index because (unlike the scenario
of separable perturbation) evolution of $\Delta \chi$ can be
separated from the spectral index of the total.

\section*{Acknowledgement}
This work is partially supported by the Grant-in-Aid for Scientific
research from the Ministry of Education, Science, Sports, and Culture,
Japan, Nos. 21111006, 22244030, 23540327 (K.K.), and 21244036 (C.M.L.).
T.M wishes to thank the organizers and the participants of COSMO13.

\appendix

\section{Running inflation model in Ref.\cite{running-inf-asym}.}
\label{app-review-EHK}
In contrast to our Multi-B model, Ref.\cite{running-inf-asym} considers
the inflaton as a source of the scale-dependent asymmetry.

In Ref.\cite{running-inf-asym}, the fraction of the total power that
comes from the curvaton has been defied as
\begin{equation}
\xi(k)=\frac{{\cal P}_{\zeta_\sigma}(k)}{{\cal
 P}_{\zeta_\sigma}(k)+{\cal P}_{\zeta_\phi}(k)},
\end{equation}
where 
\begin{eqnarray}
\label{eq-zeta-sigma}
{\cal P}_{\zeta_\sigma}(k)&=&\frac{R^2}{9}\frac{H_I^2}{\pi^2\sigma_*^2}\\
{\cal P}_{\zeta_\phi}(k)&=&\frac{G H_I^2}{\pi\epsilon_H}.
\end{eqnarray}
Note that their definitions are different from the 
component perturbations used in the conventional
curvaton scenario~\cite{curvaton-original, SVW}. 
Here we have followed the notations in Ref.\cite{running-inf-asym} and
have used $\epsilon_H\equiv -\dot{H}_I/H_I^2$ and $G\equiv 8\pi M_p^{-2}$.
$R$ is defined at the curvaton decay as
\begin{equation}
R\equiv
\frac{3\Omega_\sigma}{4\Omega_\gamma+3\Omega_\sigma+3\Omega_\mathrm{CDM}}.
\end{equation}

Their primary observation is that any scale-dependence in $\xi$ must originate
from variation in the slow-roll parameter $\epsilon_H(\phi_*)$
during inflation, which is correct only when
the evolution before the curvaton oscillation is trivial (i.e, when
$\delta \sigma/\sigma$ is constant until the beginning of the sinusoidal oscillation).

If the variation of $\epsilon_H$ is smooth, one inevitably find 
\begin{eqnarray}
\frac{d\ln \xi}{d\ln k}&=&\frac{d\ln {\cal P}_{\zeta_\sigma}}{d\ln k}
-\xi\frac{d\ln {\cal P}_{\zeta_\sigma}}{d\ln k}
-(1-\xi)\frac{d\ln {\cal P}_{\zeta_\phi}}{d\ln k}\nonumber\\
&=&(2\eta_\sigma-2\epsilon_H) -\xi(2\eta_\sigma-2\epsilon_H)\nonumber\\
&&-(1-\xi)(-4\epsilon_H+2\eta_H)\nonumber\\
&=&-(1-\xi)(2\eta_H-2\epsilon_H-2\eta_\sigma),
\end{eqnarray}
where Ref.\cite{running-inf-asym} uses the definition 
$\eta_H\equiv -\ddot{\phi}/(\dot{\phi}H_I)\simeq
M_p^2[V''(\phi)/V(\phi)]-\epsilon_H$.
Here the result is for a simple quadratic curvaton potential.

The index for the total power spectrum is
\begin{eqnarray}
n_s-1&=&\frac{d\ln\left[{\cal P}_{\xi_\sigma}+{\cal
		   P}_{\xi_\phi}\right]}{d\ln k}\nonumber\\
&=&\xi(2\eta_\sigma-2\epsilon_H)
 +(1-\xi)(2\eta_H-4\epsilon_H)\nonumber\\
&=&-2\epsilon_H +2\xi\eta_\sigma
 +(1-\xi)(2\eta_H-2\epsilon_H),
\end{eqnarray}
and the tensor to the scalar ratio is
\begin{equation}
r=16\epsilon_H(1-\xi).
\end{equation}
Therefore, we find
\begin{equation}
-\frac{d\ln \xi}{d\ln k}=n_s-1+\frac{r}{8(1-\xi)}-2\eta_\sigma.
\end{equation}
Ref.\cite{running-inf-asym} ignores $\eta_\sigma$ to find 
\begin{equation}
\label{eq-eta-not}
-\frac{d\ln \xi}{d\ln k}=n_s-1+\frac{r}{8(1-\xi)},
\end{equation}
which is wrong when $\eta_\sigma$ is significant.
Indeed, our Multi-B model considers the opposite case:
\begin{equation}
-\frac{d\ln \xi}{d\ln k}\simeq -2\eta_\sigma.
\end{equation}

They also considered a discontinuity in $\xi(k)$, which can be used to
avoid the blue spectrum.
In that case the small-scale perturbations are simply disconnected from
the large-scale anomaly.

\section{Modulated inflation}
\label{app-mod}
We introduce moduli $\chi$ and consider the action:
\begin{eqnarray}
S &=& \int d^4x \sqrt{-g} \left[
\frac{1}{4\pi G}R- \frac{1}{2}\omega(\chi)
\partial_\mu \phi \partial^\mu \phi
-\frac{1}{2} \partial_\mu \chi \partial^\mu \chi\right]\nonumber\\
&&-V(\phi)-W(\chi),
\end{eqnarray}
where $\omega(\chi)$ and the potential $W(\chi)$ are functions of the moduli.
The inflaton potential is $V(\phi)\simeq V_0$ during inflation.
The specific form of the coefficient $\omega(\chi)$ could be
$\omega(\chi)=\beta \frac{\chi^2}{M_*^2}$, where $M_*$ is a
cutoff scale.
Variation of the action leads to the equations
\begin{eqnarray}
\ddot{\phi}+3H\dot{\phi}+\frac{V'}{\omega}
+\frac{\omega'}{\omega}\dot{\phi}\dot{\chi}&=&0\\
\ddot{\chi}+3H\dot{\chi}+W'
-\frac{\omega'}{2}\dot{\phi}^2&=&0.
\end{eqnarray}
The slow-roll inflation gives
\begin{eqnarray}
\dot{\phi}&\simeq&
 -\frac{V'}{\left[3H+(\omega'/\omega)\dot{\chi}\right]\omega}\nonumber\\
&\simeq &-\frac{V'}{3H\omega},
\end{eqnarray}
where $3H\gg (\omega'/\omega)\dot{\chi}$ is assumed.

The $\delta N$ formalism gives
\begin{equation}
N_\phi=-\frac{H}{\dot{\phi}}\simeq \frac{3H^2}{V'}\omega,
\end{equation}
which leads to the mixed perturbation
\begin{equation}
 N_{\phi\chi}\simeq \frac{\omega'}{\omega}N_\phi\ne 0.
\end{equation}
For $\omega(\chi)=\beta\frac{\chi^2}{M_*^2}$, one will find
\begin{equation}
\label{eq-last}
 \frac{N_{\phi\chi}}{N_\phi}\simeq \frac{2}{\chi}.
\end{equation}


\begin{thebibliography}{1}
\bibitem{WMAPCMBA}
J.~Hoftuft, H.~K.~Eriksen, A.~J.~Banday, K.~M.~Gorski, F.~K.~Hansen and P.~B.~Lilje,
  ``Increasing evidence for hemispherical power asymmetry in the five-year WMAP data,''
  Astrophys.\ J.\  {\bf 699}, 985 (2009)
  [arXiv:0903.1229 [astro-ph.CO]].

\bibitem{PlanckCMBA}
P.~A.~R.~Ade {\it et al.}  [Planck Collaboration],
  ``Planck 2013 results. XXIII. Isotropy and Statistics of the CMB,''
  arXiv:1303.5083 [astro-ph.CO].

\bibitem{Hemisphere-original} 
  A.~L.~Erickcek, M.~Kamionkowski and S.~M.~Carroll,
  ``A Hemispherical Power Asymmetry from Inflation,''
  Phys.\ Rev.\ D {\bf 78}, 123520 (2008)
  [arXiv:0806.0377 [astro-ph]];
  A.~L.~Erickcek, S.~M.~Carroll and M.~Kamionkowski,
  ``Superhorizon Perturbations and the Cosmic Microwave Background,''
  Phys.\ Rev.\ D {\bf 78}, 083012 (2008)
  [arXiv:0808.1570 [astro-ph]].

\bibitem{Lyth-book}
  D.~H.~Lyth, A.~R.~Liddle,
  ``The primordial density perturbation: Cosmology, inflation and the
	origin of structure,'' 
  Cambridge, UK: Cambridge Univ. Pr. (2009) 497 p.

\bibitem{GZ1978}
L.~P.~Grishchuk and I.~B.~Zel'dovich, Sov. Astron. {\bf 22} (1978) 125.

\bibitem{Lyth-CMBa}
D.~H.~Lyth,
  ``The CMB asymmetry from inflation,''
  arXiv:1304.1270 [astro-ph.CO].

\bibitem{cont-ph}
  Z.~-G.~Liu, Z.~-K.~Guo and Y.~-S.~Piao,
  ``Obtaining the CMB anomalies with a bounce from the contracting phase to inflation,''
  arXiv:1304.6527 [astro-ph.CO].

\bibitem{recent-work}
 L.~Wang and A.~Mazumdar,
  ``Small non-Gaussianity and dipole asymmetry in the CMB,''
  Phys.\ Rev.\ D {\bf 88}, 023512 (2013)
  [arXiv:1304.6399 [astro-ph.CO]];
  M.~H.~Namjoo, S.~Baghram and H.~Firouzjahi,
  ``Hemispherical Asymmetry and Local non-Gaussianity: a Consistency
	Condition,''  arXiv:1305.0813 [astro-ph.CO].

\bibitem{Kanno:2013ohv}
  S.~Kanno, M.~Sasaki and T.~Tanaka,
  ``A viable explanation of the CMB dipolar statistical anisotropy,''
  arXiv:1309.1350 [astro-ph.CO].

\bibitem{Donoghue:2007ze}
 J.~F.~Donoghue, K.~Dutta and A.~Ross,
  ``Non-isotropy in the CMB power spectrum in single field inflation,''
  Phys.\ Rev.\ D {\bf 80} (2009) 023526
  [astro-ph/0703455 [ASTRO-PH]].

\bibitem{curvaton-original}
 D.~H.~Lyth and D.~Wands,
  ``Generating the curvature perturbation without an inflaton,''
  Phys.\ Lett.\ B {\bf 524}, 5 (2002)
  [hep-ph/0110002].
  T.~Moroi and T.~Takahashi,
  ``Effects of cosmological moduli fields on cosmic microwave background,''
  Phys.\ Lett.\ B {\bf 522}, 215 (2001)
  [Erratum-ibid.\ B {\bf 539}, 303 (2002)]
  [hep-ph/0110096].

\bibitem{modulateddecay}
  L.~Kofman,
  ``Probing string theory with modulated cosmological fluctuations,''
  astro-ph/0303614;
  G.~Dvali, A.~Gruzinov and M.~Zaldarriaga,
  ``Cosmological perturbations from inhomogeneous reheating, freezeout,
	and mass domination,'' 
  Phys.\ Rev.\ D {\bf 69}, 083505 (2004)
  [astro-ph/0305548];
  G.~Dvali, A.~Gruzinov and M.~Zaldarriaga,
  ``A new mechanism for generating density perturbations from inflation,''
  Phys.\ Rev.\ D {\bf 69}, 023505 (2004)
  [astro-ph/0303591].

\bibitem{IH-pt}
  T.~Matsuda,
  ``Cosmological perturbations from an inhomogeneous phase transition,''
  Class.\ Quant.\ Grav.\  {\bf 26}, 145011 (2009)
  [arXiv:0902.4283 [hep-ph]].

\bibitem{Quasar}
 C.~M.~Hirata,
  ``Constraints on cosmic hemispherical power anomalies from quasars,''
  JCAP {\bf 0909}, 011 (2009)
  [arXiv:0907.0703 [astro-ph.CO]].

\bibitem{Samuel-CMB}
  S.~Flender and S.~Hotchkiss,
  ``The small scale power asymmetry in the cosmic microwave background,''
  arXiv:1307.6069 [astro-ph.CO].

\bibitem{TiltedTurner}
 M.~S.~Turner,
  ``A Tilted universe (and other remnants of the preinflationary universe),''
  Phys.\ Rev.\ D {\bf 44}, 3737 (1991).

\bibitem{Wands:2000dp} 
  D.~Wands, K.~A.~Malik, D.~H.~Lyth and A.~R.~Liddle,
  ``A New approach to the evolution of cosmological perturbations on large scales,''
  Phys.\ Rev.\ D {\bf 62}, 043527 (2000)
  [astro-ph/0003278].

\bibitem{Dimopoulos:2003ss} 
  K.~Dimopoulos, G.~Lazarides, D.~Lyth and R.~Ruiz de Austri,
  ``Curvaton dynamics,''
  Phys.\ Rev.\ D {\bf 68}, 123515 (2003)
  [hep-ph/0308015].

\bibitem{Infcurv0}
 K.~Dimopoulos, K.~Kohri, D.~H.~Lyth and T.~Matsuda,
  ``The inflating curvaton,''
  JCAP {\bf 1203}, 022 (2012)
  [arXiv:1110.2951 [astro-ph.CO]].

\bibitem{PBHInfcurv}
K.~Kohri, C.~-M.~Lin and T.~Matsuda,
  ``PBH from the inflating curvaton,''
  arXiv:1211.2371 [hep-ph].

\bibitem{Infcurv-NonG}
S.~Enomoto, K.~Kohri and T.~Matsuda,
  ``Non-Gaussianity in the inflating curvaton,''
  Phys.\ Rev.\ D {\bf 87}, 123520 (2013)
  [arXiv:1210.7118 [hep-ph]];
  K.~Dimopoulos, K.~Kohri and T.~Matsuda,
  ``The hybrid curvaton,''
  Phys.\ Rev.\ D {\bf 85}, 123541 (2012)
  [arXiv:1201.6037 [hep-ph]].

\bibitem{modulatedinflation}
T.~Matsuda,
  ``Free light fields can change the predictions of hybrid inflation,''
  JCAP {\bf 1204}, 020 (2012)
  [arXiv:1204.0303 [hep-ph]];
 T.~Matsuda,
  ``Modulated Inflation,''
  Phys.\ Lett.\ B {\bf 665}, 338 (2008)
  [arXiv:0801.2648 [hep-ph]];

\bibitem{EKM-KLM}
S.~Enomoto, K.~Kohri and T.~Matsuda,
  ``Modulated decay in the multi-component Universe,''
  arXiv:1301.3787 [hep-ph];
K.~Kohri, C.~-M.~Lin and T.~Matsuda,
  ``Delta-N Formalism for Curvaton with Modulated Decay,''
  JCAP {\bf 1306}, 009 (2013)
  [arXiv:1303.2750 [astro-ph.CO]].

\bibitem{Modulated-curvaton}
 D.~Langlois and T.~Takahashi,
  ``Density Perturbations from Modulated Decay of the Curvaton,''
  arXiv:1301.3319 [astro-ph.CO];
  H.~Assadullahi, H.~Firouzjahi, M.~H.~Namjoo and D.~Wands,
  ``Modulated curvaton decay,''
  arXiv:1301.3439 [hep-th].

\bibitem{IH-P}
 T.~Matsuda,
  ``Generating the curvature perturbation with instant preheating,''
  JCAP {\bf 0703}, 003 (2007)
  [hep-th/0610232];
T.~Matsuda,
  ``Cosmological perturbations from inhomogeneous preheating and multi-field trapping,''
  JHEP {\bf 0707}, 035 (2007)
  [arXiv:0707.0543 [hep-th]].

\bibitem{Vilenkin-book}
A.~Vilenkin and E.~P.~S.~Shellard
  ``Cosmic Strings And Other Topological Defects,''
  Cambridge, UK: Cambridge Univ. Pr. (1994);
 A.~Vilenkin,
  ``Topological inflation,''
  Phys.\ Rev.\ Lett.\  {\bf 72}, 3137 (1994)
  [hep-th/9402085].

\bibitem{Q-ballinflation}
 T.~Matsuda,
  ``Q ball inflation,''
  Phys.\ Rev.\ D {\bf 68}, 127302 (2003)
  [hep-ph/0309339].

\bibitem{Hilltop-inflation}
 L.~Boubekeur and D.~.H.~Lyth,
  ``Hilltop inflation,''
  JCAP {\bf 0507}, 010 (2005)
  [hep-ph/0502047].

\bibitem{Hilltop-curvaton}
 T.~Matsuda,
  ``Hilltop curvatons,''
  Phys.\ Lett.\ B {\bf 659}, 783 (2008)
  [arXiv:0712.2103 [hep-ph]].

\bibitem{Topologicalcurvaton}
T.~Matsuda,
  ``Topological curvatons,''
  Phys.\ Rev.\ D {\bf 72}, 123508 (2005)
  [hep-ph/0509063];

\bibitem{refereitem}
 M.~Kawasaki, T.~Kobayashi and F.~Takahashi,
  ``Non-Gaussianity from Curvatons Revisited,''
  Phys.\ Rev.\ D {\bf 84}, 123506 (2011)
  [arXiv:1107.6011 [astro-ph.CO]];
 N.~Kobayashi, T.~Kobayashi and A.~L.~Erickcek,
  ``Rolling in the Modulated Reheating Scenario,''
  JCAP01(2014)036
  [arXiv:1308.4154 [astro-ph.CO]];
 T.~Kobayashi and T.~Takahashi,
  ``Runnings in the Curvaton,''
  JCAP {\bf 1206}, 004 (2012)
  [arXiv:1203.3011 [astro-ph.CO]].

\bibitem{running-inf-asym}
  A.~L.~Erickcek, C.~M.~Hirata and M.~Kamionkowski,
  ``A Scale-Dependent Power Asymmetry from Isocurvature Perturbations,''
  Phys.\ Rev.\ D {\bf 80}, 083507 (2009)
  [arXiv:0907.0705 [astro-ph.CO]].

\bibitem{PBHcurvaton}
  M.~Kawasaki, N.~Kitajima and T.~T.~Yanagida,
  ``Primordial black hole formation from an axion-like curvaton model,''
  Phys.\ Rev.\ D {\bf 87}, 063519 (2013)
  [arXiv:1207.2550 [hep-ph]].

\bibitem{John-tach}
J.~McDonald,
  ``Isocurvature and Curvaton Perturbations with Red Power Spectrum and Large Hemispherical Asymmetry,''
  JCAP {\bf 1307}, 043 (2013)
  [arXiv:1305.0525 [astro-ph.CO]].

\bibitem{Lyth-gfc}
  D.~H.~Lyth,
   ``Can the curvaton paradigm accommodate a low inflation scale?,''
   Phys.\ Lett.\ B {\bf 579} (2004) 239
   [hep-th/0308110].

\bibitem{John-mod} 
  J.~McDonald,
  ``Hemispherical Power Asymmetry from Scale-Dependent Modulated Reheating,''
  arXiv:1309.1122 [astro-ph.CO].

\bibitem{limit-eta}
 M.~Mijic,
  ``Particle production and classical condensates in de Sitter space,''
  Phys.\ Rev.\ D {\bf 57}, 2138 (1998)
  [gr-qc/9801094].

\bibitem{c_s-modulation}
 Y.~-F.~Cai, W.~Zhao and Y.~Zhang,
  ``CMB Power Asymmetry from Primordial Sound Speed Parameter,''
  arXiv:1307.4090 [astro-ph.CO].

\bibitem{DBI-inthesky} 
  M.~Alishahiha, E.~Silverstein and D.~Tong,
  ``DBI in the sky,''
  Phys.\ Rev.\ D {\bf 70}, 123505 (2004)
  [hep-th/0404084].
\bibitem{Modulated-nonK}
 T.~Matsuda,
  ``Modulated inflation from kinetic term,''
  JCAP {\bf 0805}, 022 (2008)
  [arXiv:0804.3268 [hep-th]].


\bibitem{SVW}
  M.~Sasaki, J.~Valiviita and D.~Wands,
  ``Non-Gaussianity of the primordial perturbation in the curvaton model,''
  Phys.\ Rev.\ D {\bf 74}, 103003 (2006)
  [astro-ph/0607627].

\end{thebibliography}
\end{document}